\documentclass[twocolumn]{aastex631} 

\usepackage{graphicx}
\usepackage{amsmath}
\usepackage{float}
\usepackage{natbib}
\usepackage{tabularx}
\usepackage{appendix}
\usepackage{longtable}
\usepackage{enumitem}
\usepackage{booktabs}
\usepackage{multirow}

\shorttitle{Forbidden BBH Formation Histories}
\shortauthors{}

\graphicspath{{./}{figures/}}

\begin{document}

\title{Forbidden Formation Histories: The Binary Black Hole Merger Rate Disfavours Long Delay Times}

\correspondingauthor{Aryanna Schiebelbein-Zwack}
\email{aryanna.schiebelbein@mail.utoronto.ca}

\author[0009-0003-3908-6112]{Aryanna Schiebelbein-Zwack}
\affiliation{David A. Dunlap Department of Astronomy and Astrophysics, University of Toronto, 50 St George St, Toronto ON M5S 3H4, Canada}
\affiliation{Canadian Institute for Theoretical Astrophysics, 60 St George St, University of Toronto, Toronto, ON M5S 3H8, Canada}

\author[0000-0002-1980-5293]{Maya Fishbach}
\affiliation{Canadian Institute for Theoretical Astrophysics, 60 St George St, University of Toronto, Toronto, ON M5S 3H8, Canada}
\affiliation{David A. Dunlap Department of Astronomy and Astrophysics, University of Toronto, 50 St George St, Toronto ON M5S 3H4, Canada}
\affiliation{Department of Physics, 60 St George St, University of Toronto, Toronto, ON M5S 3H8, Canada}

\begin{abstract}

The redshift evolution of the binary black hole (BBH) merger rate can be expressed as the convolution of the progenitor formation rate with the distribution of time delays between formation and merger. We show that starting with data-driven fits to the BBH merger rate as a function of redshift, deconvolving the inferred BBH merger rate into a delay time distribution and progenitor formation rate exposes physically incompatible delay time distributions. For a given evolution of the merger rate, certain delay time distributions are forbidden because their long-delay tails overpredict low redshift mergers independently of any assumption about the progenitor formation rate. Using delay-time distributions derived from the \texttt{COMPAS} population synthesis code in combination with the BBH merger rate inferred from GWTC-4.0, we reconstruct the physically permitted progenitor formation histories and find a steeper decline toward low redshift than the global star formation rate. We also find that the GWTC-4.0 data are in tension with formation channels that predict shallow power-law delay-time distributions ($\alpha \gtrsim -0.7$), such as stable mass transfer. Conversely, imposing the \texttt{COMPAS} predictions for the delay time distribution as a prior reduces the median merger rate inferred in GWTC-4.0 by 10\% at $z=1.5$, favoring a shallower merger rate evolution than the standard GWTC-4.0 inference. Additionally, we demonstrate that our method can constrain binary evolution physics by directly evaluating the compatibility of population synthesis parameters with gravitational wave observations. Our framework provides a model-independent avenue for ruling out regions of binary evolution and merger rate parameter space.

\end{abstract}

\section{Introduction} \label{sec:intro}

The binary black hole (BBH) merger rate is inherently informative of the progenitor stellar populations that produce merging BBHs, as it is the culmination of the birth of massive stars, the conditions of the Universe at their formation, and their subsequent evolution \citep[e.g.][]{belczynski_effect_2010,mandel_merging_2022,chruslinska_chemical_2022, vijaykumar_inferring_2023, fishbach_probing_2025}. The merger rate, $R_\text{merge}$, depends on both the rate by which progenitor stars form, $R_\text{form}$, and the delay in time until eventual merger, $\tau$ \citep[e.g.][]{2007PhR...442..166N,2010ApJ...716..615O, vitale_measuring_2019,mapelli_formation_2021, fishbach_time_2021, fishbach_ligo-virgo-kagras_2023, turbang_metallicity_2023, 2024ApJ...970..128S, 2025arXiv251006352A}. Mathematically, this is described by the linear convolution of the progenitor formation rate and the distribution of all possible delay times:
\begin{equation}\label{eq:mergerrate}
R_\text{merge}(t) = \int^{\tau_\text{max}}_{\tau_\text{min}} R_\text{form}(t-\tau)p(\tau) d\tau,
\end{equation}
where $t$ is the age of the Universe.\footnote{Throughout, we use Planck 2018 cosmology to convert between the age of the Universe $t$ and the cosmological redshift $z$ \citep{2020A&A...641A...6P}, and use cosmic time and redshift interchangeably.}
The hypothetical progenitor formation rate, $R_\text{form}$, is a modified version of the global star formation rate (SFR), assuming the observed BHs are entirely stellar remnants. These modifications account for the efficiency of BBH formation, determined by how many stars become BHs in systems that merge within a Hubble time. It remains unknown exactly which stars map to merging BBHs, and therefore gravitational wave (GW) observations are informative of the underlying physics that $R_\text{form}$ depends on. Metallicity is a key driver of BBH formation efficiency, since enriched stars produce more winds which cause radial expansion and mass loss that can prompt premature mergers or prevent BHs from forming at all \citep{belczynski_effect_2010,2020MNRAS.493L...6T,chruslinska_chemical_2022,van_son_not_2025, 2024ApJ...970..128S}. 

The distribution of delay times, $p(\tau)$, dictates the fraction of mergers that occur on short vs. long timescales post star formation. This time is dominated by the GW inspiral time \citep{1964PhDT........51P}, since the massive stars that serve as progenitors to merging systems are massive and live short lives on the order of 10 Myr. The inspiral time is dictated by the initial separations of the BHs, governed by binary evolution physics. In the context of the isolated binary formation channel, common envelope evolution is an efficient mechanism for reducing orbital energy, shrinking orbits, and leading to shorter time delays compared to stable mass transfer \citep{gallegos-garcia_binary_2021,van_son_redshift_2022, fishbach_ligo-virgo-kagras_2023, 2026A&A...706A.296K}. Alternative formation channels include chemically homogeneous evolution \citep{2016MNRAS.460.3545D, 2016MNRAS.458.2634M, 2016A&A...588A..50M, 2021MNRAS.505..663R}, triple stellar evolution \citep{2017ApJ...841...77A, 2026ApJ..1000L..59S}, dynamical evolution in star clusters \citep{2016PhRvD..93h4029R, 2019MNRAS.487.2947D, ye_redshift_2024}, or in Galactic centers with or without an active galactic nucleus (AGN) \citep{2016ApJ...831..187A,2017MNRAS.464..946S,2017ApJ...835..165B, 2018MNRAS.477.4423A, 2020A&A...638A.119G}. Altogether, the BBH merger rate as a function of redshift encodes information on star formation, stellar evolution, and environmental conditions. This makes the merger rate evolution a valuable lens through which to view stellar phylogenies and the changing landscape of the Universe.

The LIGO-VIRGO-KAGRA (LVK) collaboration \citep{2018LRR....21....3A,2015CQGra..32g4001L, 2015CQGra..32b4001A, 2021PTEP.2021eA101A} has observed the BBH merger rate evolution out to $z=1.5^{+0.2}_{-0.2}$ (90\% credibility) as of GWTC-4.0 \citep{2025arXiv250818082T,2025arXiv250818083T}. This merger rate is modeled with both strongly-modeled, or parametric, and data-driven, or non-parametric, methods. Parametric methods assume a specific underlying functional form and fit the parameters of the assumed form to the GW event data. The LVK uses a \textsc{Power-Law Redshift} model, which assumes $R_\text{merge}\propto (1+z)^\kappa$ \citep{2018ApJ...863L..41F}. The GWTC-4.0 fit resulted in $\kappa=3.2^{+0.94}_{-1.00}$ at 90\% credibility~\citep{2025arXiv250818083T}. 

Previous studies have compared strongly modelled LVK parametric merger rates to models of the progenitor formation rate and delay time distribution according to Eq.~\ref{eq:mergerrate} \citep[e.g.][]{2019MNRAS.490.3740N,2019MNRAS.482..870E,2020MNRAS.493L...6T,2022MNRAS.516.5737B,van_son_redshift_2022,turbang_metallicity_2023,2023MNRAS.523.4539K, 2024ApJ...970..128S, boesky_binary_2024}. 
These methods have provided useful insights, including establishing the metallicity dependence of BBH formation efficiency and suggesting a preference for short delay times \citep{vijaykumar_inferring_2023,turbang_metallicity_2023, 2024ApJ...970..128S}. However, approaches that utilize strongly-modeled parametric merger rates can miss features in the data not prescribed by the predetermined merger rate model structure \citep{edelman_cover_2023, callister_parameter-free_2023}. Furthermore, strong models for the progenitor formation rate and delay time distribution, even if physically motivated, may fundamentally mismatch the observed redshift evolution of the BBH merger rate~\citep{2019MNRAS.482.5012C, boesky_binary_2024, 2026arXiv260120202L, Blanchet2026}. These methods therefore always have an inherent degeneracy between the inferred delay time distribution and the progenitor formation rate; do we measure mergers from ancient stars with long time delays or recent formation with short time delays?

Alternatively, there are more flexible non-parametric approaches that fit the BBH merger rate without strong assumptions, such as the \textsc{B-spline} model used for GWTC-4.0 \citep{edelman_cover_2023, 2025arXiv250818083T}, the autoregressive BBH population inference of \citet{callister_parameter-free_2023}, {or the non-parametric delay time distributions explored in \citet{2025arXiv251006352A}.} In GWTC-4.0, the parametric \textsc{Power-Law Redshift} model is generally consistent with the flexible \textsc{B-spline} model, however with the allowance of additional structure there is a steeper decline in the merger rate from $z=0.2$ to the present and potential plateauing beyond $z=1.0$ \citep{2025arXiv250818083T}. While the lack of structure assumed \textit{a priori} in the non-parametric modeling methods can be considered a strength, this can also make interpretation of results comparatively challenging. Instead of interpreting individual hyperparameters, one must decipher coarse-grained structural features of the merger rate, which are not guaranteed to have physical validity due to the minimal constraints imposed by these models \citep{farah_things_2023}. Regardless, these {data-driven} population fits facilitate more agnostic approaches to deciphering the information embedded within the BBH merger rate than previous attempts. 

In this work, we propose a novel approach to investigate BBH origins by directly inferring the progenitor formation rate according to Eq. \ref{eq:mergerrate}. By using existing non-parametric fits to the redshift evolution of the BBH merger rate and assuming only a delay time distribution $p(\tau)$, we directly infer the progenitor formation rate without assuming its functional form. Our approach can be applied with any delay time distribution $p(\tau)$, including direct predictions from population synthesis, as long as we approximate the delay time distribution to be independent of formation redshift. In contrast to the aforementioned forward-modeling approaches, where a metallicity-specific SFR, BBH formation efficiency vs. metallicity, and delay time distribution are combined to predict a BBH merger rate for comparison to GW data, we take a backward-modeling approach (similar to \citet{2018ApJS..237....1A, fishbach_ligo-virgo-kagras_2023, 2023ApJ...950..181W}. Since the predicted merger rate can vary depending on the specific combination of population synthesis parameters and cosmological assumptions \citep[e.g.][]{2020MNRAS.493L...6T}, finding a unique match to the data is often difficult. Working backwards instead allows the data to directly dictate the structure of the formation history. This approach enables us to start with data-driven population fits and test specific assumptions one by one while flexibly fitting for the rest. By construction, the forwards inference approach restricts the model space to physically permissible results thereby obscuring potential tensions between the model and the data. Consequently, the inference may fail to capture all of the nuanced features indicated by the data \citep{Blanchet2026}. On the other hand, the backwards approach exposes nonphysical, causally excluded combinations of delay time distributions and merger rates, allowing the data to rule out certain delay time distributions independently of any assumptions about the progenitor formation rate. 

Specifically, we utilize the delay time distributions produced with the \texttt{COMPAS} population synthesis simulations \citep{2022ApJS..258...34R, van_son_redshift_2022, COMPAS:2025}, and compute the progenitor formation rate implied by their combination with the BBH merger rate evolution inferred by the \textsc{B-Spline} fit to GWTC-4.0~\citep{edelman_cover_2023,2025arXiv250818083T}. While the delay time distribution is primarily governed by binary evolution models containing many unknowns, such as mass transfer stability and supernova kick prescriptions, the progenitor formation rate represents a more macroscopic perspective, tracing the intersection of cosmic star formation history, the chemical enrichment of the Universe, and the metallicity dependence of BBH formation. We focus on computing this formation rate directly, utilizing the predicted delay time distributions derived from the intricacies of binary evolution, although this framework could be equivalently applied to compute the delay time distribution with an assumed formation history. 

We first introduce the mathematical framework for directly computing the progenitor formation rate from the BBH merger rate in Section \ref{sec:fourier}. In Section \ref{sec:rform}, we use the \texttt{COMPAS} delay time distributions to compute the BBH progenitor formation rate. We also introduce the potential for physical incompatibilities between delay time distributions and merger rate combinations. We explore how this incompatibility can constrain models and population synthesis parameters in Section \ref{sec:constraint}. Finally, we conclude in Section \ref{sec:conclusion}.

\section{Deconvolving the Merger Rate}\label{sec:fourier}
The BBH merger rate can be used in combination with a delay time distribution to infer the underlying formation rate (or vice-versa). Although it is possible to solve for $R_\text{form}$ by rearranging Eq. \ref{eq:mergerrate} directly in the time domain \citep{fishbach_probing_2025}, such operations (e.g., recursive deconvolution) are computationally expensive. Instead, we utilize the convolution theorem, which states that a convolution in the time domain is a simple product in the frequency domain \citep{arfken1985mathematical}. Thus, Eq. \ref{eq:mergerrate} is simplified by transforming from the time domain to the frequency domain via the Fourier transform:

\begin{equation}\label{eq:mergerrate_fourier}
\mathcal{F}\{R_\text{merge}\} = \mathcal{F}\{R_\text{form}\} \cdot  \mathcal{F}\{p(\tau)\} 
\end{equation}

This allows for a non-parametric reconstruction of the progenitor formation rate, in contrast to previous approaches that perform a joint fit of formation and time delays, constraining the formation rate to specific parameterized models, and potentially resulting in poor fits to the LVK data. Rearranging Eq. \ref{eq:mergerrate_fourier} gives the progenitor formation rate:

\begin{equation}\label{eq:prograte}
 R_\text{form}= \mathcal{F}^{-1} \left\{ \frac{\mathcal{F}\{R_\text{merge}\}}{\mathcal{F}\{p(\tau)\}} \right\}
\end{equation}

Unlike the traditional forward modeling approaches, our framework reduces the assumptions necessary to characterize the BBH progenitor population, allowing the data to directly dictate the structure of the formation history. 

\subsection{Computational Methodology}

The numerical computation of the formation rate via deconvolution (Eq. \ref{eq:prograte}) is not as straightforward as the analytical expression suggests. We first transform the theoretical time delay distributions and the merger rate posterior curves into the frequency domain via the Fast Fourier Transform (FFT) \footnote{Our code is available at \url{https://github.com/arysch/Deconvolving_merger_rates}}. The FFT assumes a periodic signal and computes a circular convolution, which can cause the end of the signal to wrap-around and contaminate the beginning. Thus, to recover the physically meaningful linear convolution of Eq. \ref{eq:mergerrate}, and prevent the high redshift data from contaminating the low redshift regime, we pad the ends of the inputs with zeros to provide a sufficient buffer for any wrap-around artifacts.

Furthermore, the bounded nature of the LVK observation horizon is also a source of spectral leakage, since the edges of the data act as discontinuities. In the transformation to the frequency domain, these sharp boundaries produce noise in the form of oscillatory artifacts, known as the Gibbs phenomenon caused by the attempt to replicate the jump with finite sinusoids \citep{arfken1985mathematical}. To mitigate these effects, we apply smooth tapering to the edges of the data. For the merger rate in the local Universe, $z<0.1$, we blend the merger rate to $\mathcal{R}_\text{merge} = 0$ using a cosine window. 

At $z > 1.5$, where the BBH merger rate is unconstrained with GWTC-4.0 data, we impose a boundary condition on the past merger rate. The BBH merger rate has contributions from both recently-formed systems that have short delay times and old systems that formed much earlier but had long delay times. 
Thus, any assumption for $\mathcal{R}_\text{merge}(z>1.5)$ implies additional formation at high redshift, which propagates through long delay times to the merger rate at low redshifts. This correspondingly reduces the formation rate required at intermediate times to match the observed BBH mergers. Accordingly, the assumption that no mergers occur beyond the current LVK horizon provides a conservative upper limit on the formation rate in the observed era. This assumption would attribute all of the observed mergers to formation within (or very shortly before) the detection horizon. Conversely, assuming a higher merger rate in the past naturally requires formation at higher redshifts, and these ancient stars contribute a tail of long time delay mergers into the observation window, thereby reducing the rate of new formation required in order to match the LVK data. 

To explore the impact of this assumption, we consider a range of boundary conditions in addition to the limiting case of $\mathcal{R}_\text{merge}(z>1.5) = 0$. We also consider constant asymptotic merger rates of $\mathcal{R}_\text{merge}(z > 1.5) \in \{100, 200, 300, 400, 500\} \, \text{Gpc}^{-3}\text{yr}^{-1}$. Again, to prevent sharp jumps in $\mathcal{R}_\text{merge}(z)$, we blend the GWTC-4.0 posterior curves to these asymptotic values by $z=1.55$ using a cosine taper, ensuring adequate suppression of the edge-driven numerical noise while maximizing the retention of the data.

The formation rate reconstruction is performed using Wiener deconvolution, which modifies the division of Eq. \ref{eq:prograte} for numerical stability \citep{press2007numerical}:

\begin{equation}
\mathcal{F}\{R_\text{form}\} = \mathcal{F}\{R_\text{merge}\} \cdot \frac{\mathcal{F}\{p(\tau)\}^*}{|\mathcal{F}\{p(\tau)\}|^2 + \lambda}
\end{equation}

Instead of dividing by the Fourier transform of the delay time distribution, which can be problematic when $\mathcal{F}\{p(\tau)\}\rightarrow 0$, we implement the equivalent approach of multiplying by the complex conjugate normalized by its squared magnitude. With the denominator transformed into a real number, as opposed to the complex form it was in before, we can add a term for regularization. The regularization term, $\lambda$, dominates when $|\mathcal{F}\{p(\tau)\}|^2\rightarrow 0$ and prevents numerical instabilities. In essence, $\lambda$ suppresses contributions from components that are poorly constrained and would otherwise introduce spurious fluctuations in the solution. We adopt $\lambda = 10^{-3} \text{max}(|\mathcal{F}\{p(\tau)\}|^2)$, a fixed fraction of the peak power of the Fourier transform of the delay time distribution. This smooths features with power below 0.1\% of the peak.

Finally, the resulting formation rate is normalized by the temporal step size, $\Delta t$. This ensures that the discrete summation performed by the FFT remains physically consistent with the original continuous integral definition in Eq. \ref{eq:mergerrate}. This step is necessary to preserve the physical units of $\text{Gpc}^{-3} \text{yr}^{-1}$ and to ensure the total integrated number of events is conserved.

We have confirmed the validity of this procedure by testing that the computed formation rate, when convolved with the delay time distribution, reproduces the original merger rate posterior curve (see Fig. \ref{fig:quota}).

\section{Reconstructing the Progenitor Formation Rate}\label{sec:rform}

We utilize the deconvolution method presented in Section \ref{sec:fourier} to infer $\mathcal{R}_\text{form}(z)$ from posterior draws of $\mathcal{R}_\text{merge}(z)$, as inferred by the GWTC-4.0 \textsc{B-Spline} fit, in combination with delay time distributions obtained with the population synthesis code \texttt{COMPAS} \citep{2022ApJS..258...34R, van_son_no_2022,fishbach_2023_10019887, COMPAS:2025}. The suite of delay time distributions contain variations on mass transfer, supernova, and remnant mass settings. Thus, our analysis accounts for the uncertainties inherent in binary stellar evolution physics. Unless otherwise stated, we adopt the conservative assumption that $\mathcal{R}_\text{merge}(z>1.5) = 0$, as explained in Section \ref{sec:fourier}.

\begin{figure}
    \centering
    \includegraphics[width=1\linewidth]{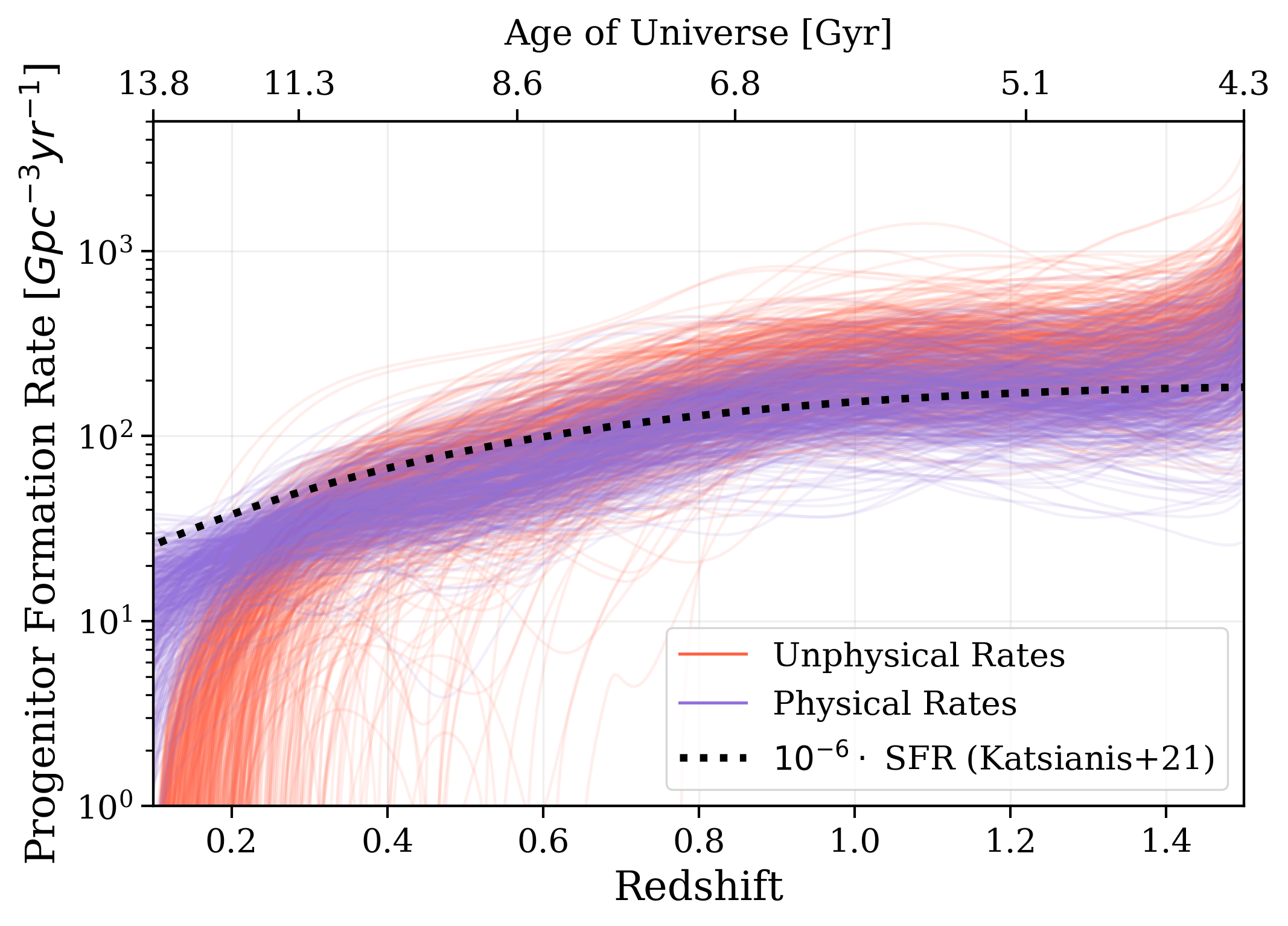}
    \caption{Random draws of the computed BBH progenitor formation rate as a function of cosmic time (or equivalently, redshift), $\mathcal{R}_\text{form}(t)$, obtained by deconvolving the GWTC-4.0 BBH merger rate posterior draws, $\mathcal{R}_\text{merge}(t)$, and delay time distributions, $p(\tau)$, from \texttt{COMPAS}. The formation rates that remain positive, and are therefore physical, throughout the observable redshift range are shown in purple, and those that drop to negative values and are nonphysical are shown in orange. A rescaled SFR is shown for comparison by the dotted black line. The progenitor formation rate $\mathcal{R}_\text{form}$ declines more rapidly than the global SFR at late times.}
    \label{fig:formationrates}
\end{figure}

The reconstructed progenitor formation rates are shown in Fig. \ref{fig:formationrates}, with the purple curves showing physically-consistent rates and orange showing nonphysical rates (as we return to shortly). A scaled version of the SFR is also depicted, with a benchmark efficiency of BBH formation of $10^{-6} \,\text{M}_\odot^{-1}$ \citep{fishbach_ligo-virgo-kagras_2023}. Assuming a range of delay time distributions from \texttt{COMPAS}, we infer that the progenitor formation rate approximately follows the SFR as reported by \citet{katsianis_observed_2021}, but appears to decline more rapidly than the SFR with cosmic time. This is consistent with the results of \citet{Blanchet2026}, and is expected due to the metallicity dependence of BBH formation. As the Universe becomes more metal enriched, BBH formation decreases \citep{belczynski_effect_2010}.

\begin{figure*}
    \centering
    \includegraphics[width=1\linewidth]{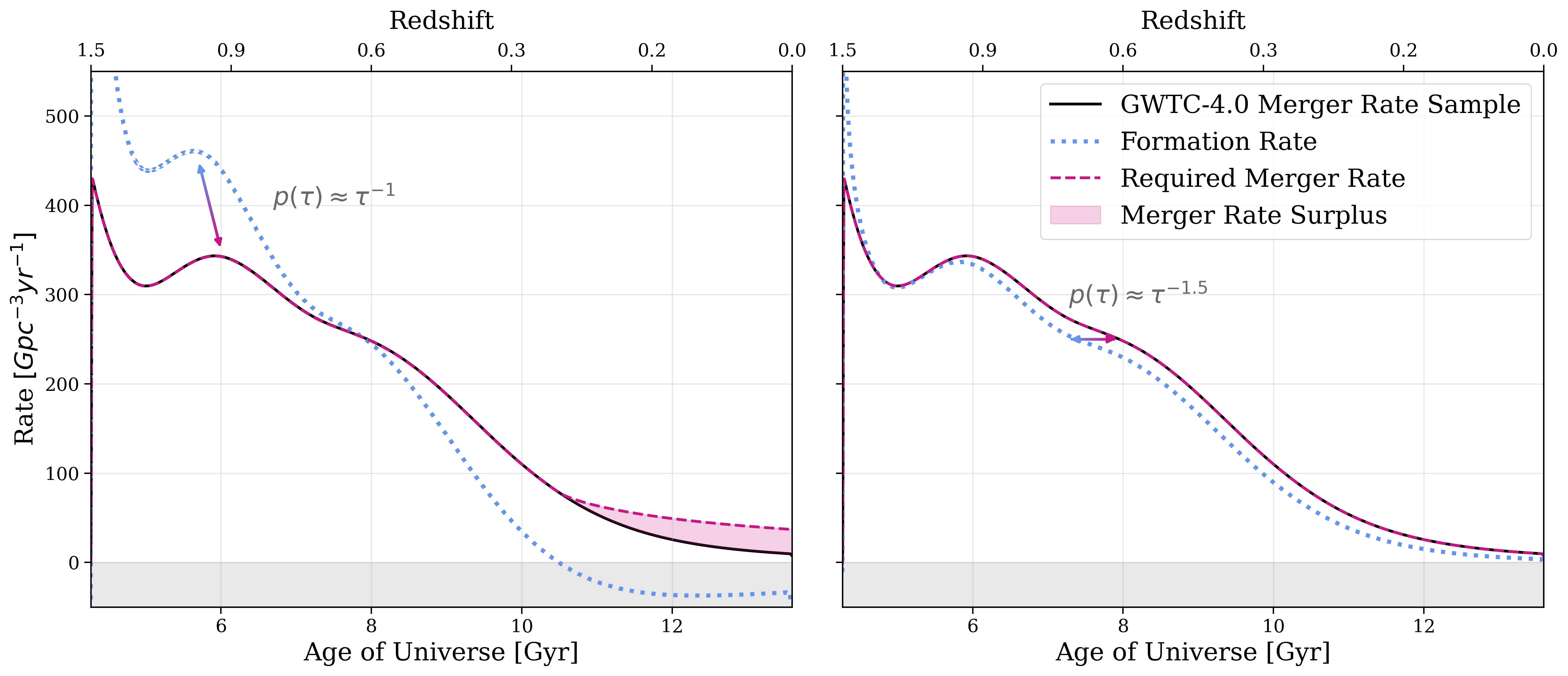}
    \caption{An example GWTC-4.0 merger rate posterior curve (black) compared with the reconstructed formation rate (blue dotted) for $p(\tau)\propto \tau^{-1}$ (\textit{left}) and $p(\tau)\propto \tau^{-1.5}$ (\textit{right}). The dashed pink line shows the merger rate obtained by reconvolving the physically allowed (non-negative) formation rate with the specified time delay distribution. In other words, this is the merger rate implied by the cumulative contribution of past formation weighted by the delay-time distribution. The shallower time delay distribution ($\alpha=-1$) requires a merger rate that exceeds the observed merger rate at late times, meaning that the combination of the the past formation rate and long delay times predict more mergers than are observed. Consequently, the formation rate becomes negative to force the required merger rate to match the observed merger rate. In contrast, for the steeper delay time distribution ($\alpha=-1.5$), which favours shorter delays, the recomputed merger rate remains consistent with the data, and the formation rate stays non-negative. }
    \label{fig:quota}
\end{figure*}

Interestingly, there are combinations of $\mathcal{R}_\text{merge}$ and $p(\tau)$ that are unphysical; they require $\mathcal{R}_\text{form}$ to become negative at late times (the orange curves in Fig. \ref{fig:formationrates}). This occurs because any formation event at high redshift creates a merger quota for all subsequent times due to the long-delay time tail of $p(\tau)$. If the assumed $p(\tau)$ has a particularly high fraction of long delays, the stars formed at, for instance, $z=1$, may account for more mergers than are actually observed at $z=0.5$. To resolve this surplus and match the observed LVK merger rate, the formation rate is forced to ``remove" stars that would be merging with long delay times. This phenomena is exhibited in Fig. \ref{fig:quota}. Put simply, these models are unphysical because the observed merger rate drops more steeply than the intrinsic ``memory" of the delay-time distribution allows. 
Because a broad delay time distribution smooths out the BBH merger rate, steeply evolving merger rates are incompatible with broad delay time distributions regardless of the assumed progenitor formation rate.

The emergence of this unphysical behaviour is an opportunity; we can directly test the viability of models based on the compatibility of their delay time distributions with the GW data. 

\section{Physical Constraints}\label{sec:constraint}
Here, we translate the appearance of negative formation rates into constraints on the components of Eq.~\ref{eq:prograte}, namely the \texttt{COMPAS} delay time distributions (Section~\ref{subsec:ptau}), the LVK merger rate measurements (Section~\ref{subsec:mergerrates}), and the joint parameter space describing both quantities (Section~\ref{subsec:joint}). 

\subsection{Population Synthesis Delay Time Distributions}\label{subsec:ptau}

We explore the compatibility of different delay time distributions, $p(\tau)$,  with the data-driven merger rates, $\mathcal{R}_\text{merge}(t)$, inferred in GWTC-4.0. We start by parameterizing the delay time distribution as a power law in order to gain intuition about the combinations of delay time distributions and merger rate curves that are physically compatible. The power law form for $p(\tau)$ is a common choice, motivated by the fact that both the initial binary separation and the subsequent GW inspiral time follow power laws \citep{mandel_merging_2022}:

\begin{equation}\label{eq:ptau}
    p(\tau) = \frac{\alpha+1}{\tau_{\text{max}} ^{\alpha+1}-\tau_{\text{min}} ^{\alpha+1}} \tau^{\alpha}
\end{equation}
where we fix $\tau_\text{min}$ to 10 Myr, the approximate lifetime of a massive star, and $\tau_\text{max}$ to the age of the Universe. The power-law index, $\alpha$, determines the steepness of the delay time distribution in log-space and consequently the fraction of mergers that are expected to occur at short or long delay times. A more negative $\alpha$ leads to a steeper power-law, a preference for shorter time delays, and therefore corresponds to closer initial separations for BBHs. For instance, common envelope evolution has been attributed to $\alpha \approx -1$, whereas stable mass transfer which is less efficient at removing orbital energy is characterized with $\alpha \approx -0.35$ \citep{fishbach_ligo-virgo-kagras_2023}. Hence, it is possible to constrain which binary stellar evolution models are preferred depending on if their delay times are compatible with the GW data. 

We quantify the agreement of each of the \texttt{COMPAS} delay time distributions with the GWTC-4.0 \textsc{B-Spline} merger rate posterior curves.
For reference, we also compute the best-fitting power-law approximation to each $p(\tau)$ from the \texttt{COMPAS} suite, characterized by the slope $\alpha$.
The compatibility of each $p(\tau)$ is given by the fraction $\mathcal{P}$ of $\mathcal{R}_\text{merge}$ posterior draws that produce a strictly positive $\mathcal{R}_\text{form}$ over the redshift range $z=[0.1, 1.5]$: 
\begin{equation}
\mathcal{P}(\mathcal{R}_\text{form}>0 \ \forall \ z \in [0.1,1.5] \ | \  \alpha,\text{O4a})
\end{equation}

We truncate our range at $z=0.1$ instead of $z=0$ because the uncertainties in the merger rate estimate increase due to the rareness of local detections, with the prior on the \textsc{B-Spline} model driving the inference at $z < 0.1$  \citep{edelman_cover_2023}.

\begin{figure}
    \centering
    \includegraphics[width=1\linewidth]{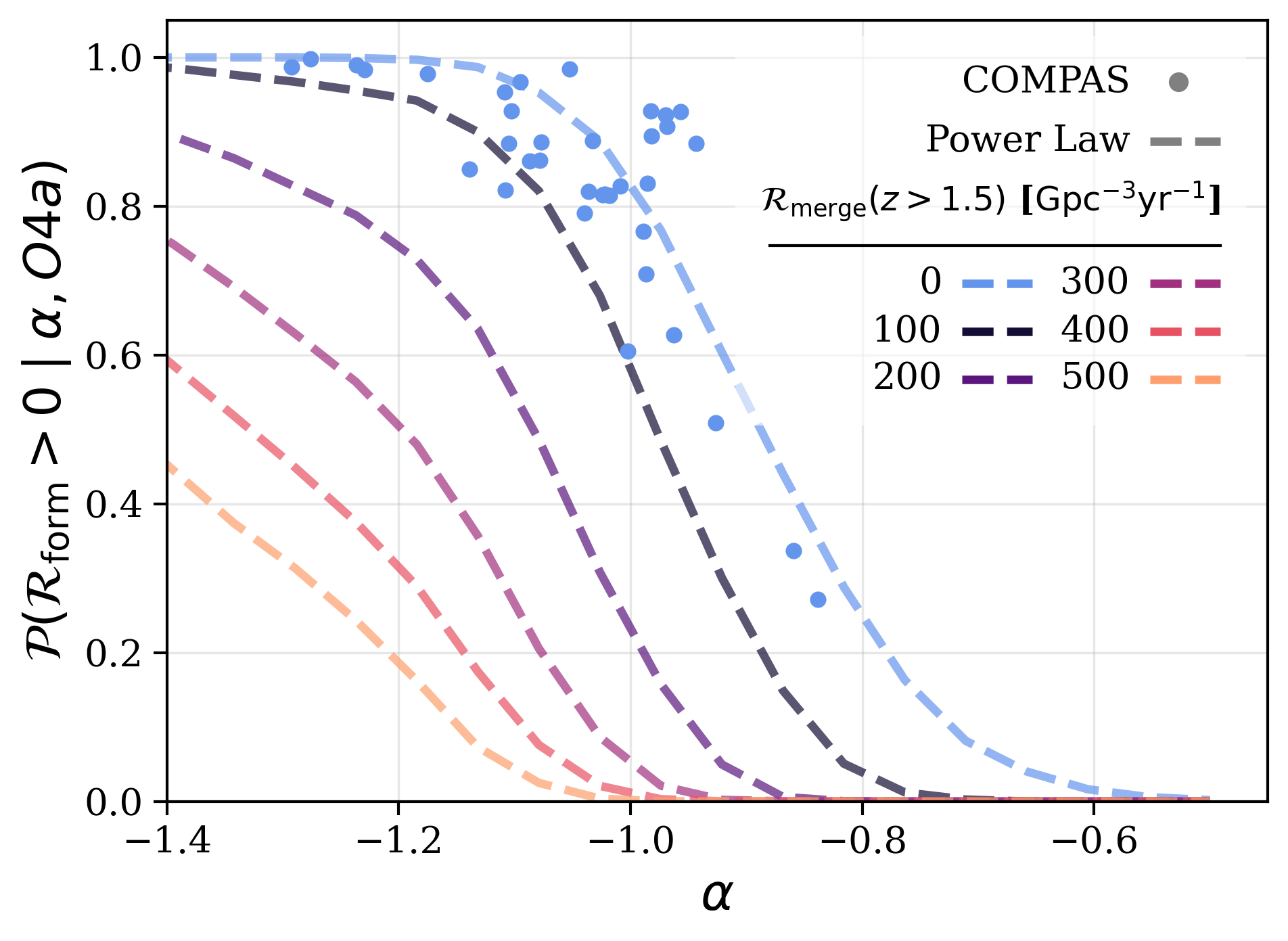}
    \caption{The fraction of the GWTC-4.0 merger rate posterior $\mathcal{P}$ that produces physical, positive formation rates with delay time distributions characterized here by their best fit power-law index, $\alpha$. The blue dots represent the individual time delay distributions computed with a suite of population synthesis simulations from \texttt{COMPAS}. The dashed lines show the compatibility of idealized power-law time delay distributions with the GWTC-4.0 inferred merger rate evolution, under varying assumptions for the unobserved merger rate at $z > 1.5$. Steeper distributions (more negative $\alpha$), which prioritize shorter delay times, have larger $\mathcal{P}$ and are thus preferred by the observed GW data. Assuming a higher merger rate beyond the detection horizon reduces the compatibility of a given delay time distribution, as the merger ``quota" from past formation increasingly exceeds the local observed rate.}
    \label{fig:passrate_alpha}
\end{figure}

As seen in Fig. \ref{fig:passrate_alpha}, the \texttt{COMPAS} delay time distributions that are steeper and prefer shorter delays ($\alpha<-1.2$) are consistent with nearly all of the GWTC-4.0 merger rate posterior draws ($\mathcal{P} \approx 1$). The shallowest delay time distributions, with $\alpha \approx -0.85$, are only compatible with $\approx 30\%$ of the GWTC-4.0 merger rate draws. Note, these delay time distributions are global, representing the combined contributions of stable mass transfer and common envelope evolution. Separating these distributions into distinct evolutionary components reveals potentially shallower indices; for instance, stable mass transfer alone is estimated to follow $\alpha \approx -0.35$ \citep{fishbach_ligo-virgo-kagras_2023}.

We additionally determine $\mathcal{P}(R_\text{form}>0 \ \forall \ z \in [0.1,1.5] \ | \  \alpha,\text{O4a})$ using a pure power-law delay time distribution, also shown in Fig. \ref{fig:passrate_alpha}. The \texttt{COMPAS} simulated results follow the trend of the pure power-law distribution, albeit with more scatter in the compatibility fraction $\mathcal{P}$ due to deviations from the perfect power-law. The power-law time delay distribution agrees with 95\% of the GWTC-4.0 merger rate posterior when $\alpha<-1.08$ and excludes all but 5\% when $\alpha >-0.67$. In other words, $\alpha<-1.08$ is preferred by a factor of 19 compared to $\alpha >-0.67$. 

Recall, the formation rates we compute, and therefore the compatibility fraction $\mathcal{P}$, are upper bounds, due to the assumption of $\mathcal{R}_\text{merge}(z>1.5)=0$. However, mergers that occur beyond the detection horizon will decrease the formation rate amplitudes shown in Fig. \ref{fig:formationrates}. As a consequence, even more combinations of $\mathcal{R}_\text{merge}$ and $p(\tau)$ will be incompatible and $\mathcal{P}(R_\text{form}>0 \ \forall \ z \in [0.1,1.5] \ | \  \alpha,\text{O4a})$ will decrease for a fixed $\alpha$ in Fig. \ref{fig:passrate_alpha}. The compatibility of the GWTC-4.0 $\mathcal{R}_\mathrm{merge}(z)$ posterior draws with different power-law delay time distributions, assuming various constant boundary conditions $\mathcal{R}_\text{merge}(z>1.5)$, is also shown in Fig. \ref{fig:passrate_alpha}. For increased past merger rates, $\alpha$ must be comparatively more negative to produce a non-negative formation rate. If $\mathcal{R}_\text{merge}(z>1.5)=300 \text{ Gpc}^{-3}\text{year}^{-1}$, we require $\alpha<-1.73$ in order to agree with 95\% of the GWTC-4.0 merger rate posterior, and 95\% of the posterior is excluded when $\alpha > -0.99$.

These results are aligned with previous work that found GWTC-2 \citep{fishbach_time_2021} and GWTC-3 data preferred short time delays (e.g., $\alpha \leq -1.32$, \citealp{2024ApJ...970..128S}; $\alpha \leq -1.55$, \citealp{vijaykumar_inferring_2023}; $\alpha \leq -1$, \citealp{2023MNRAS.523.4539K}). Not only do we confirm this result using the latest GW data in GWTC-4.0, we find that short delay times are preferred independently of any assumptions about the progenitor formation rate (past work either fixed the progenitor formation rate or marginalized over a fixed range of assumptions). This observational finding appears to be in tension with claims that stable mass transfer evolution can account for the majority of BBH mergers\citep{2019MNRAS.490.3740N, gallegos-garcia_binary_2021, van_son_no_2022}. Stable mass transfer produces wide binaries that have long time delays between star formation and merger \citep[$\alpha \approx -0.35$;][]{fishbach_ligo-virgo-kagras_2023, 2026A&A...706A.296K}, in conflict with the GW data that favours steeper delay time distributions. Other channels that also predict long delay times include chemically homogeneous evolution \citep[at $Z\approx 0.2Z_\odot$,][]{2016MNRAS.458.2634M}, triple stellar evolution \citep{2017ApJ...841...77A, 2026ApJ..1000L..59S}, and high efficiency common envelope evolution \citep{2021A&A...647A.153B, van_son_redshift_2022}.

\begin{figure}
    \centering
    \includegraphics[width=1\linewidth]{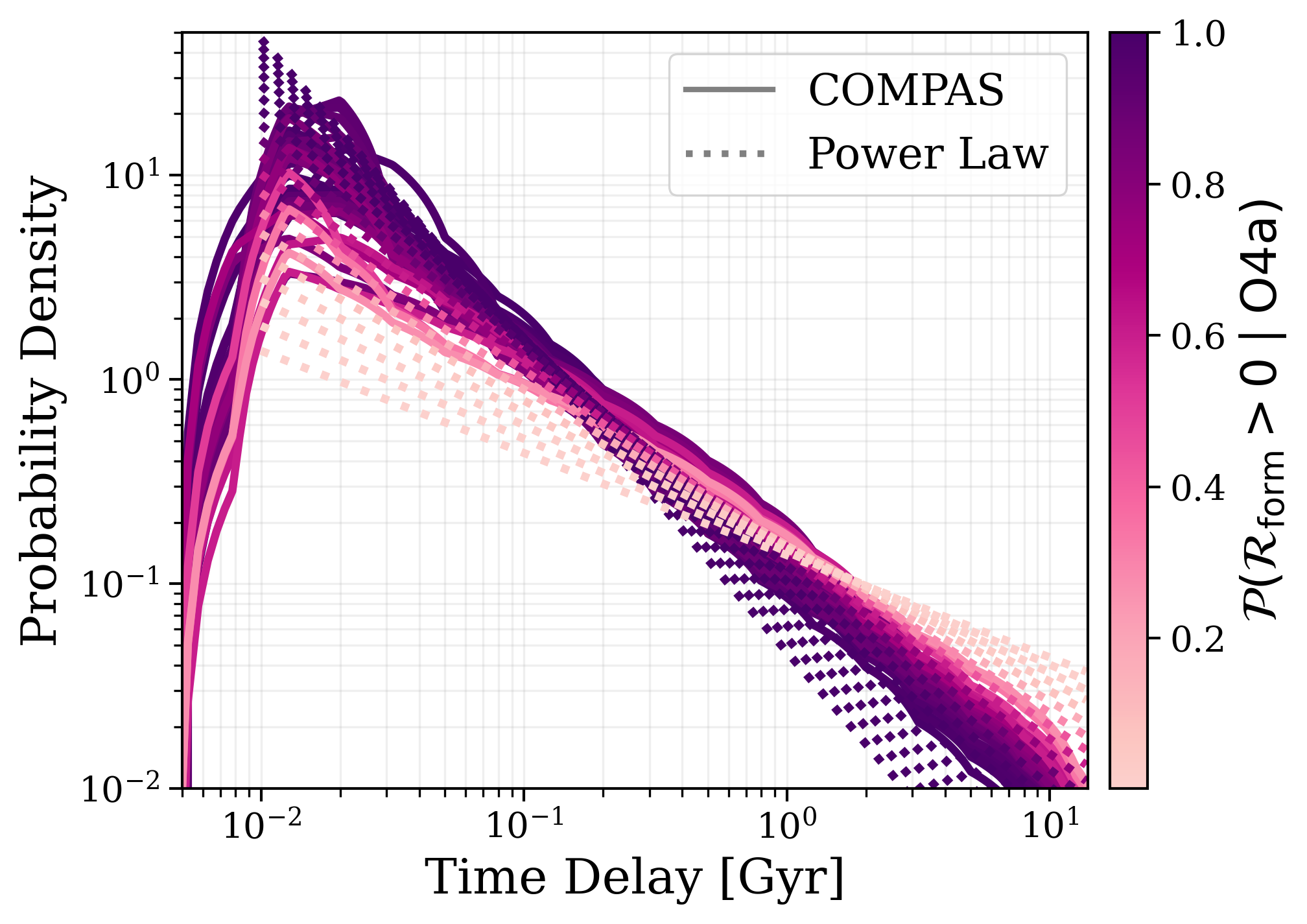}
    \caption{The distribution of delay times from \texttt{COMPAS} simulations (solid) and analytic power laws (dotted). The colour of each curve represents the fraction of the GWTC-4.0 merger rate posterior that is consistent with a physical formation history under the given delay time distribution. The steeper the delay time distribution, and the higher the probability density at short delay times, the higher the compatibility with GWTC-4.0.}
    \label{fig:ptau}
\end{figure}

The \texttt{COMPAS} and analytic power-law delay time distributions are shown in Fig. \ref{fig:ptau}, with colours representative of the compatibility fraction $\mathcal{P}$. Visually, steeper distributions with higher probability density at shorter time delays are favoured; i.e., they are more likely to produce a physical formation rate when deconvolved with the inferred GWTC-4.0 merger rate. Conversely, distributions with significant support at long time delays are less compatible with the data.

In principle, we can translate the compatibility of different delay time distributions into constraints on the binary population synthesis models. The binary populations synthesis parameter space is multi-dimensional with complex, non-linear interactions. {While a comprehensive investigation of these parameters utilizing our backward-modeling deconvolution to identify forbidden formation histories is reserved for future studies (building upon the multi-dimensional inference of, e.g., \citet{2018MNRAS.477.4685B}), we perform a targeted exploration of a single parameter to demonstrate how our methodology can be applied to constrain stellar evolution physics.}

\begin{figure}
    \centering
    \includegraphics[width=1\linewidth]{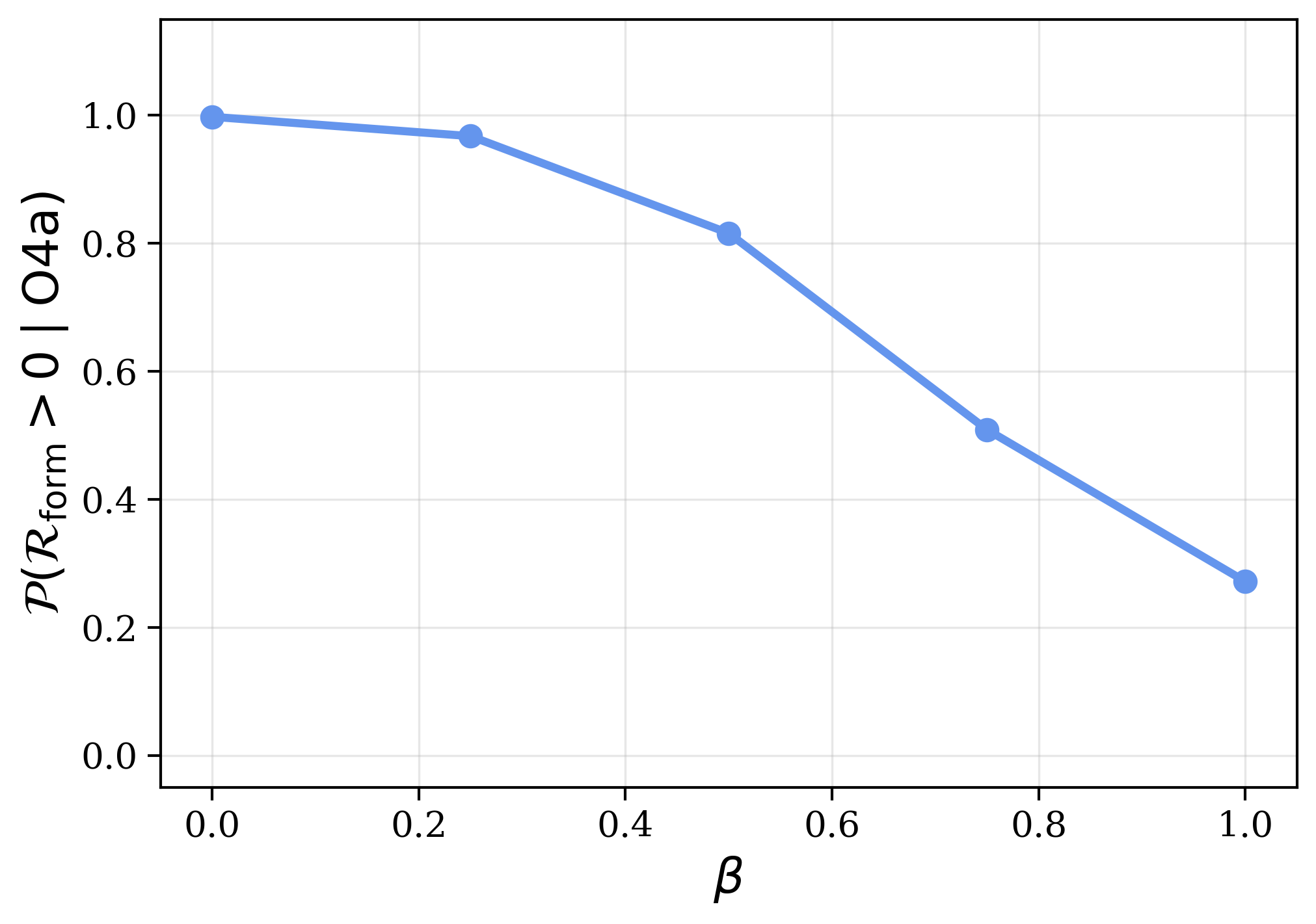}
    \caption{The compatibility of various assumptions for the \texttt{COMPAS} mass transfer efficiency parameter $\beta$ with the GWTC-4.0 data. As $\beta$, the fraction of mass successfully accreted during mass transfer, increases, the compatibility decreases. In other words, GWTC-4.0 disfavours large values of $\beta$. This is because larger $\beta$ corresponds to higher retention of angular momentum, a higher orbital separation, and longer time delays.}
    \label{fig:popsynth}
\end{figure}

The \texttt{COMPAS} parameter $\beta$ represents the mass transfer efficiency, defined as the fraction of transferred mass that is successfully accreted by a companion star \citep{2022ApJS..258...34R, van_son_no_2022}. A smaller $\beta$ implies that a corresponding fraction of the transferred mass is lost from the system, carrying away angular momentum and resulting in orbital shrinkage (depending on the mass ratio of the system). In contrast, a larger $\beta$ corresponds to conservative mass transfer where the mass and angular momentum are retained in the system. This prevents dramatic orbital contraction, leading to wider binary separations and consequently longer time delays. We show the compatibility of the inferred merger rates from GWTC-4.0 with \texttt{COMPAS} delay time distributions of varied $\beta$ (fixing the other parameters to their default values) in Fig. \ref{fig:popsynth}.

The compatibility is strongest for lower $\beta$ values and steadily declines as $\beta$ increases. Since $\beta=0$ corresponds to maximal mass loss and the most efficient orbital shrinkage, these models produce the short delay times that have been shown to be more consistent with the GW data. On the other hand, higher $\beta$ values are in greater tension with the data, as they produce an excess of long delay time systems that are unable to recover the steeply evolving merger rate at low redshift without requiring unphysical formation histories. We find that $\beta = 1$ is disfavoured by a factor of $\approx3.7$ compared to $\beta = 0$.

This analysis provides a proof-of-concept rather than a complete picture of the $\beta$ values consistent with data. We have neglected potential degeneracies with other parameters, such as the core mass fraction and the donor star radial response to mass loss, among many others. Furthermore, a rigorous analysis would require testing compatibility across other observational dimensions, such as the BH mass and spin distributions, to ensure that the models favored by the merger rate evolution also are consistent with the other properties of the astrophysical BBH population. {Additionally, the merger rate of compact objects of different masses may probe metallicity evolution, star formation rate densities, and binary evolution parameters to varied degrees and will need to be considered in tandem to break degeneracies \citep{2020MNRAS.493L...6T}.}

\subsection{Binary Black Hole Merger Rates}\label{subsec:mergerrates}
\begin{figure}
    \centering
    \includegraphics[width=1\linewidth]{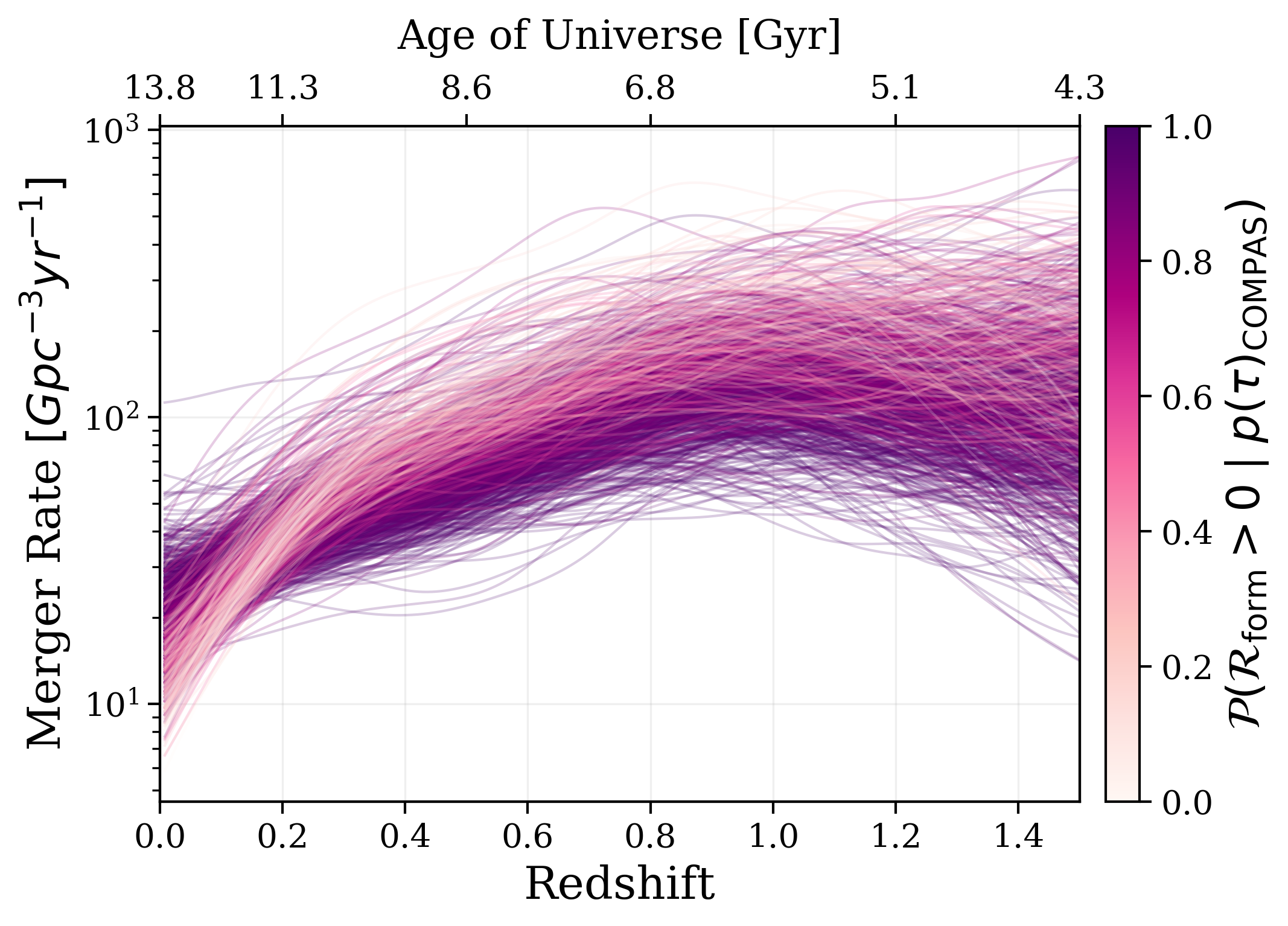}
    \caption{Posterior samples of the BBH merger rate inferred under the \textsc{B-Spline} model, coloured by their compatibility with the \texttt{COMPAS} delay time distributions. The disfavored curves (i.e., those that correspond to a physical formation rate for a smaller fraction of $p(\tau)$ draws from the \texttt{COMPAS} suite) have larger merger rates at high redshift ($z\sim 1.5$) and correspondingly sharper rate declines towards late times. The \texttt{COMPAS} delay time predictions thus prefer merger rates that have shallower evolution across redshift.}
    \label{fig:mergerrate}
\end{figure}

Having evaluated the compatibility of various delay time distributions with the GW data, we now invert our focus to analyze the compatibility of the BBH merger rate posterior with the assumed time delays. Fig. \ref{fig:mergerrate} illustrates the posterior distribution of the GWTC-4.0 BBH merger rates inferred under the \textsc{B-Spline} model. As we established previously, a merger rate that declines too quickly cannot be reconciled with the ``memory" inherent to delay-time distributions. Thus, the merger rate samples with the lowest compatibility with the \texttt{COMPAS} simulations are those exhibiting the sharpest slopes. In contrast, the curves most consistent with the delay-time simulations are those that have a flatter structure across cosmic time. 

If most BBH mergers come from isolated binary evolution and current understanding of binary stellar evolution physics is indeed correct, then the physical requirement of a positive formation rate (Fig.~\ref{fig:quota}) further constrains the merger rate posterior. The 90\% credible upper limit of the BBH merger rate at $z=1.5$ shifts from $\mathcal{R}_\text{merge} = 334 \text{ Gpc}^{-3}\text{year}^{-1}$ under the flexible \textsc{B-Spline} inference down to $\mathcal{R}_\text{merge} = 299 \text{ Gpc}^{-3}\text{year}^{-1}$ when we condition on \texttt{COMPAS} delay time distributions. (Again, this is independent of any assumption about the progenitor formation rate other than that it remains positive at all redshifts.) While the numerical shift is modest ($\approx  10\%$) using the conservative assumption $\mathcal{R}_\text{merge}(z>1.5) = 0$, it represents a systematic downward revision of the peak merger rate driven solely by the prior requirement that $\mathcal{R}_\text{form} >= 0$. Thus, while non-parametric, data-driven fits to the GW data afford maximum flexibility, they by definition do not enforce compatibility with any physical assumptions and may allow for BBH merger histories that are fundamentally inconsistent with star formation in the Universe.

Recent findings by \citet{farah_steep_2026} indicate that the redshift evolution of hierarchically merged systems may be steeper than the total BBH population, and correspondingly, the subpopulation of BBH mergers from isolated binary evolution may evolve less steeply with redshift. Therefore, segregating the isolated binary contribution to the total merger rate may improve the agreement with population synthesis predictions.

\subsection{Joint Constraints on the Delay Time Distribution and Merger Rate: Power Law Approximations}\label{subsec:joint}

In the preceding sections, we demonstrated that certain combinations of BBH merger rates and delay time distributions are incompatible, requiring negative, and thus unphysical, progenitor formation rates. We now map the theoretical boundaries of this incompatibility.

As mentioned, the merger rate and time delay distribution are commonly approximated as power laws. The merger rate is parameterized in redshift as $R_\text{merge}\propto (1+z)^\kappa$, where $\kappa=3.2^{+0.94}_{-1.00}$ in the latest GWTC-4.0 results \citep{2025arXiv250818083T}. The time delay distribution is parameterized in time by Eq. \ref{eq:ptau}, where $\alpha$ is a consequence of the underlying stellar binary evolution physics. We explore the deconvolution (Eq. \ref{eq:prograte}) of these two theoretical parameterizations to determine the physically permitted regions of the $\alpha$--$\kappa$ parameter space. 

For a given merger rate index $\kappa$, there exists a critical delay time distribution index $\alpha_\text{crit}$ for which any shallower delay time distribution ($\alpha>\alpha_\text{crit}$) results in unphysical, negative formation rates. This occurs because a shallow delay time distribution acts as a smoothing kernel, as shown in Fig. \ref{fig:quota} and explained in Section \ref{sec:rform}. 

\begin{figure}
    \centering
    \includegraphics[width=1\linewidth]{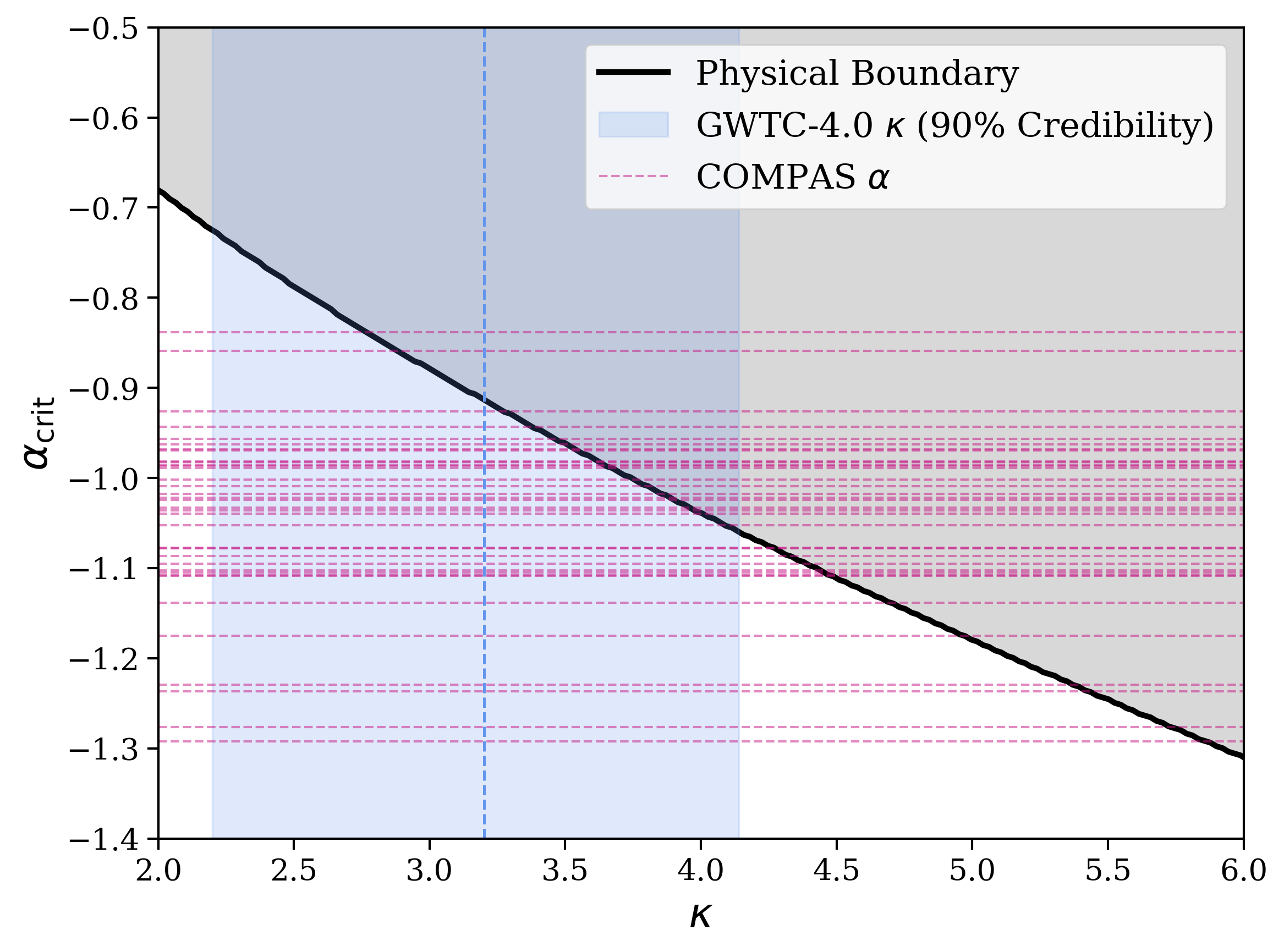}
    \caption{The critical time delay distribution index $\alpha$ as a function of merger rate evolution index $\kappa$. For a specific $\kappa$, the delay time index $\alpha$ is constrained to be below the black boundary curve. The region above the curve is physically forbidden, as all $\alpha$--$\kappa$ combinations here result in negative progenitor formation rates. The blue shaded band illustrates the GWTC-4.0 90\% credible interval for $\kappa$, as well as the median as a dashed line ($\kappa$ = $3.2^{+0.94}_{-1.00}$). The pink lines indicate the $\alpha$ values fit to the \texttt{COMPAS} delay time distributions. Some of the \texttt{COMPAS} binary evolution models are in tension with the observed cosmic merger history.}
    \label{fig:alpha_kappa}
\end{figure}

Fig. \ref{fig:alpha_kappa} illustrates the relationship $\alpha_\text{crit}(\kappa)$. The shaded region above the curve denotes the forbidden area of parameter space, while the area below the boundary is physically permitted. Current LVK observations indicate $\alpha \lessapprox -0.73$ by the 90\% credible interval. Future observing runs will yield more events and expand the redshift horizon, thus tightening this limit. Notably, at the median observed $\kappa\approx 3.2$, several \texttt{COMPAS} variations are already ruled out, suggesting that current binary evolution models may produce delay times that are too long to be consistent with the observed cosmic BBH merger history.

These results, derived from idealized power-law formulations, are broadly consistent with the analysis of the \texttt{COMPAS} delay-time distributions and \textsc{B-Spline} merger rates presented in Section \ref{subsec:ptau}. For instance, that analysis shows that 95\% of the merger rate posterior is excluded for indices $\alpha > -0.67$, compared to $\alpha>-0.73$ here. As indicated by the \texttt{COMPAS} data points in Fig. \ref{fig:passrate_alpha}, distributions with similar $\alpha$ values can exhibit a spread in compatibility. This variance is due to the flexibility in those models, which reveal localized structure in the merger rate that can provide more leverage for model rejection. For instance, if the merger rate exhibits a particularly steep decline over a narrow redshift range it can rule out delay time distributions that a simple smooth global power law might otherwise permit. Thus, while the $\alpha$--$\kappa$ plane provides a clear and interpretable theoretical boundary for the population, the full non-parametric treatment presented before captures the nuanced wiggles in the cosmic history that are most sensitive to unphysicality. 

\section{Conclusion}\label{sec:conclusion}

In this work, we have explored the consequences of the fact that the features observed in the BBH merger rate are inherited from the underlying progenitor formation rate and the delay-time distribution. By viewing the merger rate as a convolution of these two physical functions, we presented a method in Section \ref{sec:fourier} for recovering a data-driven formation rate, assuming an arbitrary delay time distribution (or vice versa). By pairing the non-parametric \textsc{B-Spline} merger rates inferred in GWTC-4.0 and the delay time distributions from a suite of \texttt{COMPAS} population synthesis simulations, we reduce the systematic biases inherent to forward modeling rigidly parameterized functions. Avoiding the strict \textit{a priori} structural restrictions of parameterized models is vital, as these directly dictate the formation histories we seek to reconstruct.

We find that the BBH progenitor formation rate declines more rapidly than the shape of the SFR with increasing cosmic time (decreasing redshift; Fig. \ref{fig:formationrates}). Our analysis also revealed that there are combinations of observed merger rates and delay time distributions that are incompatible, as they would imply unphysical, negative formation rates (see Fig. \ref{fig:quota}). This means that we can constrain physical models independently of any assumption about the progenitor formation rate. 

Given the \textsc{B-Spline} fit the merger rate evolution, the GW data are more compatible with delay time distributions that indicate a preference for short delay times (Fig. \ref{fig:passrate_alpha} \& Fig. \ref{fig:ptau}). Approximating the delay time distribution $p(\tau)$ as a power law, we find that the GWTC-4.0 data prefer a power-law index $\alpha<-1.08$ by a factor of 19 over $\alpha>-0.67$. The favouring of short delay times appears to be at odds with binary evolution models where stable mass transfer is the primary formation channel \citep{gallegos-garcia_binary_2021,van_son_no_2022}. However, because our analysis treats the BBH population globally, the apparent disconnect may be driven by contamination from subpopulations that do not evolve from isolated binary evolution. For example, dynamically-assembled hierarchical mergers may exhibit more rapid redshift evolution~\citep{farah_steep_2026}. 
Future studies that isolate the isolated binary evolution subpopulations may yet find coherence with a larger fraction of the population synthesis predictions. 

As a proof of concept, we explore the physical compatibility of the population synthesis parameter, $\beta$, with GWTC-4.0. The simulations with lower values of $\beta$ showed greater compatibility with the GWTC-4.0 merger rate inference, suggesting a preference for non-conservative mass transfer (Fig. \ref{fig:popsynth}). Although a rigorous analysis incorporating the degeneracies between different parameters of population synthesis models is left to future studies, this serves as a demonstration of the usefulness of our presented methodology.

If we ``update" the BBH merger rate posterior, as illustrated in Fig. \ref{fig:mergerrate}, we determine that the merger rates in agreement with the \texttt{COMPAS} delay time distributions are shallower in structure. Conditioning on adherence to the \texttt{COMPAS} delay time distributions lowers the median BBH merger rate by 10\% to $\mathcal{R}_\text{merge}(z=1.5) = 299 \text{ Gpc}^{-3} \text{ year}^{-1}$. This shift demonstrates that physical consistency requirements can act as an informative prior. If the majority of BBH mergers can be attributed to \texttt{COMPAS}-like binary evolution, then the merger rate evolution must be shallower than the data alone might suggest. Recall that by assuming $\mathcal{R}_{\text{merge}}(z > 1.5) = 0$, we provide a conservative bound on the progenitor formation rate. However, as future observations reveal the merger rate at high redshift, the corresponding high-$z$ formation will account for an increasing fraction of local observations due to long delay time systems. This increases the physical pressure on the deconvolution; if a delay time distribution predicts more `inherited' mergers than are observed, that model will be rejected. Consequently, as the high-$z$ merger rate is uncovered, the space of physically consistent delay time distributions and local merger rates will be increasingly restricted.

Lastly, using theoretical power-law models for both the BBH merger rate, $\mathcal{R}_\mathrm{merge} \propto (1+z)^\kappa$, and delay time distribution, $p(\tau) \propto \tau^\alpha$, we mapped out the forbidden region of $\alpha$ -- $\kappa$ parameter space. This provides a physically motivated boundary for the inferred LVK merger rates; no global delay time distribution shallower than the critical curve can produce the observed merger rate evolution. This constraint can be made regardless of any assumption about the progenitor formation rate. The 90\% credible interval for $\kappa$ in GWTC-4.0 excludes power-law delay time distributions with $\alpha \gtrapprox -0.7$.

By imposing the physical constraint of a positive formation rate, we provide a framework for making robust and model-independent constraints on stellar binary evolution. This approach offers a significant advantage over jointly forward modeling the delay time distribution and progenitor formation rate, as it bypasses the need for high-dimensional priors on star formation, metallicity evolution, and their degeneracies with the delay time distribution. {Conversely, this framework can be inverted to test whether an inferred delay time distribution is physically admissible, providing a complementary consistency check to the causality constraints described in \citet{2025arXiv251006352A}.}

As future data sets expand our observation horizons, the constraints presented in this work will only become stronger. Next generation observatories will detect merging BBHs as far back in time as they exist, allowing for a more complete picture of the metallicity-dependent progenitor formation history \citep{2010CQGra..27s4002P,2012CQGra..29l4013S,2019BAAS...51g..35R,2020JCAP...03..050M,2024CQGra..41x5001G,2024A&A...681A..56S,fishbach_probing_2025}. This longitudinal view of the merger rate will provide us with a deep `fossil record' of stellar phylogenies, providing a powerful independent probe of cosmic metallicity and massive star formation over a Hubble time. By simply rejecting models that require unphysical progenitor histories, we will significantly restrict the permissible parameter space for binary evolution, narrowing down our understanding of how BBHs form in our Universe.

\section*{Acknowledgments}

We thank Asad Hussain and Christopher Berry for comments on our initial manuscript, as well as Will Farr, Martyna Chruslinska, Lieke van Son, Reed Essick, Aditya Vijaykumar, and Amanda Farah for insightful discussions and helpful exchanges. 
ASZ expresses appreciation to Jonathan Zhang, Michael Grehan, and Tanisha Ghosal for their helpful feedback on the visual presentation of the figures in this work and also acknowledges the support of the Natural Sciences and Engineering Research Council of Canada - Canada Graduate Scholarships - Doctoral (NSERC-CGS-D) program. 
MF acknowledges support from the Natural Sciences and Engineering Research Council of Canada (NSERC) under grant RGPIN-2023-05511, the Alfred P. Sloan Foundation, and the Ontario Early Researcher Award ER24-18-170.
This material is based upon work supported by NSF’s
LIGO Laboratory which is a major facility fully funded
by the National Science Foundation.
This research was supported in part by grant NSF PHY-2309135 to the Kavli Institute for Theoretical Physics (KITP).

\software{ 
NumPy \citep{harris2020array}, SciPy \citep{virtanen2020scipy}, Matplotlib \citep{hunter2007matplotlib}, pandas \citep{mckinney2010data}, Seaborn \citep{waskom2021seaborn}, JAX \citep{jax2018github}, NumPyro \citep{2019arXiv191211554P},  Astropy \citep{astropy2013, astropy2018, astropy2022}, and corner.py \citep{foreman2016corner}.}




\bibliography{bibtex}{}
\bibliographystyle{aasjournal}

\end{document}